\documentclass[bibyear,letter]{aa}
\usepackage{graphicx}
\usepackage[varg]{txfonts}
\usepackage{placeins}
\begin{document}

\title{
  Imaging the innermost circumstellar environment of the 
  red supergiant WOH G64 in the Large Magellanic Cloud
\thanks{
  Based on observations collected at the European Southern Observatory under
  ESO programmes 106.219D.001, 106.219D.002, 110.23RT.004,
  097.D-0605(A), and 71.B-0558(A) as well as at the Rapid Eye Mount
  Telescope under the Chilean National Time Allocation Committee programme
  CN2024A-73.
}
}
%\subtitle{
%}

\author{K.~Ohnaka\inst{1} 
\and
K.-H.~Hofmann\inst{2} 
\and
G.~Weigelt\inst{2} 
\and
J. Th. van Loon\inst{3}
\and
D.~Schertl\inst{2}
\and
S. R. Goldman\inst{4}
}

\offprints{K.~Ohnaka}

\institute{
  Instituto de Astrof\'isica, 
  Departamento de Ciencias F\'isicas, 
  Facultad de Ciencias Exactas,
  Universidad Andr\'es Bello, 
  Fern\'andez Concha 700, Las Condes, Santiago, Chile\\
\email{k1.ohnaka@gmail.com}
\and
Max-Planck-Institut f\"{u}r Radioastronomie, 
Auf dem H\"{u}gel 69, 53121 Bonn, Germany
\and
Lennard-Jones School of Chemical and Physical Sciences, 
Keele University, Staffordshire
ST5 5BG, U.K.
\and
Space Telescope Science Institute, 3700 San Martin Drive, Baltimore,
MD 21218, U.S.A.
}

\date{Received / Accepted }

\abstract
% Context
{
  Significant mass loss in the red supergiant (RSG) phase has great influence
  on the evolution of massive stars and their final fate as supernovae. 
}
% Aim
{
  We present near-infrared interferometric imaging of the 
  circumstellar environment of the dust-enshrouded RSG WOH G64 in the Large
  Magellanic Cloud.

}
% Methods
{
  WOH G64 was observed with the GRAVITY
  instrument at ESO's Very Large Telescope Interferometer (VLTI) at
  2.0--2.45~\mbox{$\mu$m}. 
  We succeeded in imaging the innermost circumstellar environment of
  WOH G64 -- the first interferometric imaging of an RSG outside the Milky
  Way. 
}
% Results
{
  The reconstructed
  image reveals elongated compact emission with a semimajor and semiminor
  axis of $\sim$2 and $\sim$1.5~mas ($\sim$13 and 9~\mbox{$R_{\star}$}),
  respectively. 
  The GRAVITY data show that 
  the stellar flux contribution at 2.2~\mbox{$\mu$m}\ at the time of our 
  observations in 2020 is much lower than predicted by 
  the optically and geometrically thick
  dust torus model based on the VLTI/MIDI data taken in 2005 and 2007.
  We found a significant change in the near-infrared spectrum of WOH G64:
  while the (spectro)photometric data taken at 1--2.5~\mbox{$\mu$m}\ before 2003
  show the spectrum of the central RSG with \mbox{H$_2$O}\ absorption,
  the spectra
  and {\it JHK$^{\prime}$} photometric data
  taken after 2016 are characterized by a
  monotonically rising continuum 
  with very weak signatures of \mbox{H$_2$O}. This spectral change likely took place
  between December 2009 and 2016. On the other hand, the mid-infrared
  spectrum obtained in 2022 with VLT/VISIR agrees well with the spectra
  obtained before 2007.
  }
% Conclusions
{
  The compact emission imaged with GRAVITY and the near-infrared spectral
  change suggest the formation of hot new dust close to the star,
  which gives rise to the
  monotonically
  rising near-infrared continuum and
  the high obscuration of the central star. The elongation of the emission
  may be due to the presence of a bipolar outflow or
  effects of an unseen companion.
}

   %  END OF ABSTRACT

\titlerunning{Near-infrared interferometric imaging of the circumstellar
  environment of the LMC RSG WOH G64}
\authorrunning{Ohnaka et al.}
\maketitle

\begin{figure*}
\sidecaption
\rotatebox{0}{\includegraphics[width=12cm]{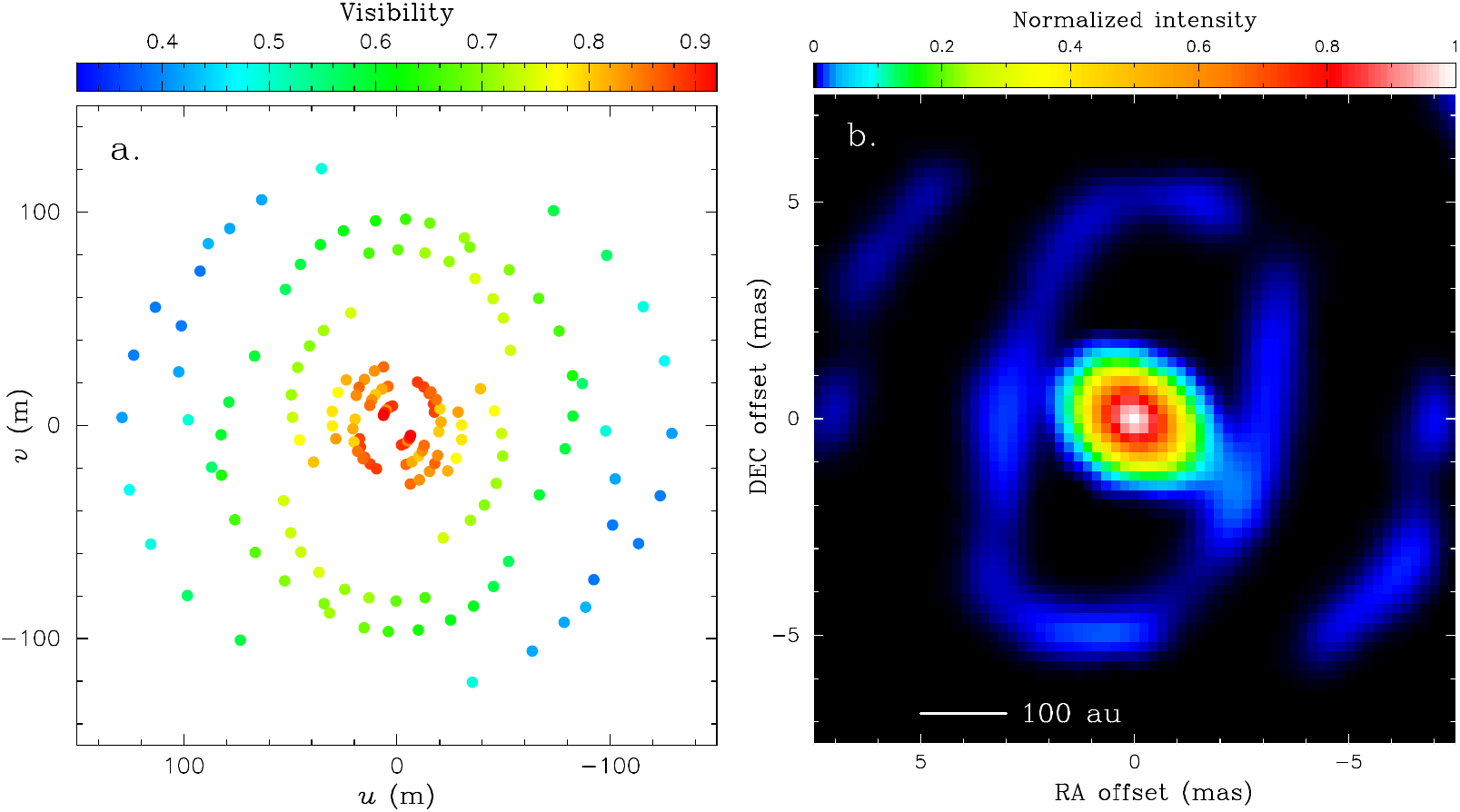}}
%\end{center}
\caption{
  Visibility and image of WOH G64 obtained from our VLTI/GRAVITY 
  observations. 
  {\bf a:} $u\varv$ coverage of our VLTI/GRAVITY observations of WOH G64
  with the calibrated visibility color-coded. North is up, east is to the left.
  {\bf b:} Image of WOH G64 reconstructed at 2.2~\mbox{$\mu$m}\ (with a spectral
  window of 0.2~\mbox{$\mu$m}) using IRBis with the maximum entropy 
  regularization. North is up, east is to the left. 
}
\label{uv_image}
\end{figure*}

\section{Introduction}
\label{sect_intro}

Significant mass loss in the red supergiant (RSG) phase is of great importance 
for the evolution of massive stars before they end their life in a supernova 
(SN) explosion.
The RSGs at advanced evolutionary stages experience drastic mass loss with 
a mass-loss rate as high as $\sim \!\! 10^{-4}$ \mbox{$M_{\sun}$~yr$^{-1}$}\ 
(Goldman et al. \cite{goldman17}).
%\LEt{***Avoid beginning a sentence with an acronym (especially a paragraph),
%  and do not begin with a numeral, formula, or symbol.} 
Recent analyses of very early-phase spectra of SNe -- taken within a 
day after the explosion -- suggest significant 
increases in the mass-loss rate in the RSG phase 
before the SN explosion (e.g., Yaron et al. \cite{yaron17}; 
Moriya et al. \cite{moriya18}; Zhang et al. \cite{zhang23}). 

High-spatial-resolution observations of some 
RSGs reveal salient deviations from spherical symmetry in 
their circumstellar environment 
(e.g., Wittkowski et al. \cite{wittkowski98}; Monnier et al. \cite{monnier04}; 
Humphreys et al. \cite{humphreys07}; O'Gorman et al. \cite{ogorman15}). 
Nonspherical mass loss can also be exemplarily seen in 
the dust ring around SN1987A, which is considered to have been shed 
in the RSG phase before the progenitor evolved into a blue supergiant
and exploded (Crotts \& Heathcoat \cite{crotts91}).
Given the high multiplicity rate among massive stars
(Mason et al. \cite{mason09}; Sana et al. \cite{sana12}), 
the asymmetric, enhanced mass loss in the RSG phase, which can be
driven by binary interaction (e.g., Ercolino et al. \cite{ercolino24};
Landri \& Pejcha \cite{landri24}), 
is essential not only for
better understanding the evolution of massive stars 
but also for interpreting early-phase SN spectra.

The RSGs in the Large Magellanic Cloud (LMC) have the great advantage that their 
distances are much better known (50 kpc, Pietrzy\'nski et al.
\cite{pietrzynski13}) compared to those of their Galactic counterparts.
WOH G64 is the brightest RSG in the mid-infrared in the LMC, exhibiting
a huge infrared excess with a high mass-loss rate on
the order of $10^{-4}$~\mbox{$M_{\sun}$~yr$^{-1}$}\ (Goldman et al. \cite{goldman17}).
For this reason, it has been a subject of multiwavelength studies 
from the visible to the radio (e.g., van Loon et al. \cite{vanloon96};
Levesque et al. \cite{levesque09}; Matsuura et al. \cite{matsuura16}). 
Ohnaka et al. (\cite{ohnaka08}) succeeded in 
spatially resolving the circumstellar dust environment of WOH G64 
using the mid-infrared interferometric instrument MIDI at the Very Large Telescope
Interferometer (VLTI). 
Their 2D radiative transfer modeling shows that the observed spectral energy distribution and the
visibilities measured at 8--13~\mbox{$\mu$m}\  
can be simultaneously reproduced by a geometrically and optically thick torus 
viewed nearly pole-on.
Moreover, the luminosity of $\sim \! 2.8 \times 10^5$~\mbox{$L_{\sun}$}\ 
derived from the dust torus model brought WOH G64 to
fair agreement with the evolutionary track with an initial mass of 25~\mbox{$M_{\sun}$}.

However, because MIDI was a two-telescope interferometer, it was not possible
to obtain an image of WOH G64. In this Letter, we present the first 
infrared interferometric imaging of WOH G64.

\section{Observations and data reduction}
\label{sect_obs}

We carried out interferometric observations of WOH G64 with GRAVITY
(GRAVITY Collaboration \cite{gravity17}) at VLTI on December 15 and
26, 2020 (UTC), at 2.0--2.45~\mbox{$\mu$m}, using the Auxiliary Telescope (AT)
configurations A0-G1-J2-J3 and A0-B2-C1-D0 
with a maximum projected baseline length of 129~m 
(Program ID: 106.219D.001/002, P.I.: K.~Ohnaka). 
We also observed HD32956 (F0/2IV/V, uniform-disk diameter = 0.17~mas,
JMMC catalog: Bourges et al. \cite{bourges17})
and HD37379 (F6/7V, uniform-disk diameter = 0.16~mas)
for the interferometric and spectroscopic calibration. 
Our GRAVITY observations are summarized in Table~\ref{obslog}. 

To increase the signal-to-noise (S/N) of
the results, the raw GRAVITY data taken with a spectral resolution of 500
were first spectrally binned using a running box car filter, which
resulted in a spectral resolution of 330. The spectral binning was applied
to both the science target and the calibrators as well as the raw calibration
files needed to create the P2VM.  
The spectrally binned raw data were then reduced with the GRAVITY pipeline
ver~1.4.1\footnote{https://www.eso.org/sci/software/pipelines/gravity/}.
The errors in the calibrated visibilities of WOH G64 were computed from
the errors given by the pipeline and the variations in the transfer
function calculated from all the calibrators observed on each night. 

Figure~\ref{uv_image}a shows the $u\varv$ coverage of our GRAVITY observations
of WOH G64 with the visibility color-coded. The visibility falls off more
rapidly in the northeast-southwest direction than in the northwest-southeast direction, which suggests
that the object appears larger in the northeast-southwest direction.
  As Figs.~\ref{viscpplotL} and \ref{viscpplotS} show, 
  the visibilities and closure phases show 
  no trace of the CO bands, although the 2.3~$\mu$m band head is weakly 
  seen in the spectrum (Fig.~\ref{wohg64nirspec}b). 
We reconstructed images from the GRAVITY data using IRBis
(Hofmann et al. \cite{hofmann14}) and MiRA
(Thi\'ebaut \cite{thiebaut08}).
IRBis selects the best reconstruction based on the fit to the measurements
and the distribution of the positive and negative residuals between the
measured quantities (visibilities and closure phases) and those of the 
reconstructed image. 
The best reconstruction with IRBis was obtained with the 
maximum entropy regularization using the cost functions 1 and 2 defined in Hofmann et al. (\cite{hofmann14}).
For the reconstruction with MiRA, two different regularizations
(pixel difference quadratic and pixel intensity quadratic)
%\footnote{They are called qsmooth and quadratic in MiRA, respectively.})
were employed,
and the optimal value of the
hyperparameter was determined with the L-curve method.
The spectrum of WOH G64 extracted from the GRAVITY data was spectroscopically
calibrated, as is described in Appendix~\ref{appendix_gravity_spec}. 
%The reconstructed images were flux-calibrated using the absolute flux 
%of the photometrically calibrated spectrum. 

We also obtained $J$-, $H$-, and $K^{\prime}$-band photometric data on August 11, 2024, with the REMIR camera at the Rapid Eye Mount (REM)
Telescope\footnote{http://www.rem.inaf.it} (Molinari \cite{molinari19}). 
Details of the REM observations are described in
Appendix~\ref{appendix_rem}.

\begin{figure*}
\begin{center}
\resizebox{0.99\hsize}{!}{\rotatebox{0}{\includegraphics{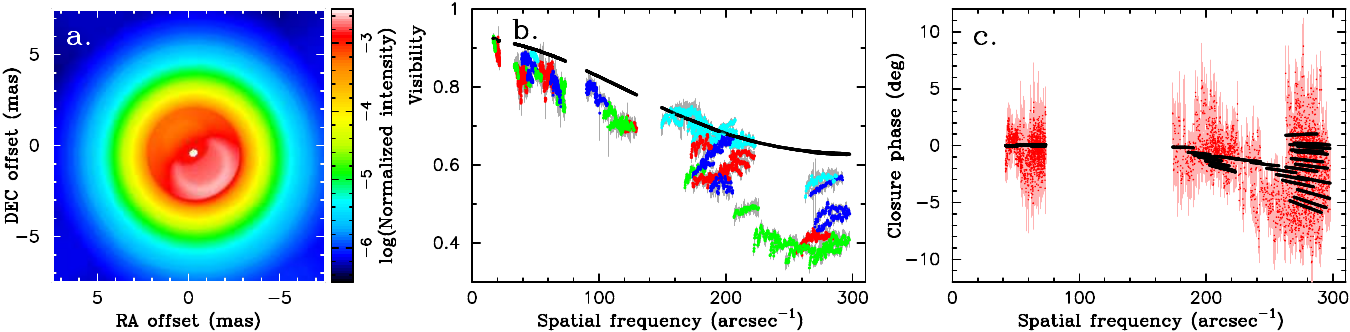}}}
%\resizebox{17.6cm}{!}{\rotatebox{0}{\includegraphics{image_viscpplot_PAcolor.pdf}}}
\end{center}
\caption{
  Comparison of the dust torus model of Ohnaka et al. (\cite{ohnaka08}) with
  the GRAVITY data.
  {\bf a:} Image predicted by the dust torus model at 2.2~\mbox{$\mu$m}\
  with an inclination angle of 20\degr. Note that the central star's
  intensity (white dot at the center) is at least $\sim$400 times
  higher than that of the torus. North is up, east is to the left.
  {\bf b:} Comparison of the visibility between the dust torus model and
  the GRAVITY data.
  The black dots (they appear to be solid lines due to the high density) 
  represent the 2.2~\mbox{$\mu$m}\ visibilities
  predicted by the dust torus model.
    The dots with the error bars represent the GRAVITY data with the
    position angle (PA) of the projected baseline color-coded:
    red (0\degr$\le$PA$<$45\degr), green (45\degr$\le$PA$<$90\degr),
    blue (90\degr$\le$PA$<$135\degr), and
    light blue (135\degr$\le$PA$<$180\degr). 
  {\bf c:} Comparison of the closure phase between the dust torus model and
  the GRAVITY data. The model closure phases are shown in black, while the
  measurements are plotted with the red dots with the error bars.
}
\label{model2008}
\end{figure*}

\section{Results}
\label{sect_res}

Figure~\ref{uv_image}b shows the image reconstructed at 2.2~\mbox{$\mu$m}\
with \mbox{IRBis} using the data between 2.1 and 2.3~\mbox{$\mu$m}. 
The resolution of the image is $\sim$1~mas. 
Figure~\ref{gravity_viscpplot} shows comparisons of the measured visibilities
and closure phases with those computed from the \mbox{IRBis} reconstructed
image. 
The image reconstructed with MiRA is shown in Fig.~\ref{wohg64images},
in which only the image obtained with the pixel difference quadratic
regularization is presented, because the result obtained with the pixel
intensity quadratic regularization is similar.

Both images obtained with IRBis and MiRA reveal compact elongated emission
with a major and minor axis of $\sim$4~mas and 3~mas, respectively,
with a position angle of the major axis of $\sim$56\degr.
This corresponds to the position angle dependence of the visibility
seen in Fig.~\ref{uv_image}a. 
The stellar radius of 1730~\mbox{$R_{\sun}$}\ derived by Ohnaka et al. (\cite{ohnaka08})
and the distance of 50~kpc translate into a stellar angular radius of
0.16~mas. Therefore, the semimajor and semiminor axes of the elongated emission
correspond to $\sim$13 and $\sim$9~\mbox{$R_{\star}$}, respectively. 
The central star does not clearly appear as a point source in our
reconstructed image. 
This could be because the central star is very faint or because it is entirely obscured.
The visibilities measured at the longest baselines tend to be flat 
as a function of spatial frequency 
(Figs.~\ref{model2008}b and ~\ref{gravity_viscpplot}a),
which can be explained by the presence of a 
point source. However, data at even longer baselines are needed to
confirm this and derive the fractional flux contribution of the central
star in the $K$ band. 

A faint elliptical ring-like structure ($\la$3\% of the peak intensity) with
a semimajor and semiminor axis of $\sim$5~mas and 3~mas 
($\sim$31 and 19~\mbox{$R_{\star}$}), respectively, can be seen in dark blue in
both images reconstructed with IRBis and MiRA.
This ring-like structure might be interpreted as the inner rim of a dust disk
or torus viewed from an intermediate inclination angle, higher than in
the torus model of Ohnaka et al. (\cite{ohnaka08}). 
The ring's appearance in the images reconstructed with different algorithms
(MiRA and IRBis) lends support to it being real. 
However, as Fig.~\ref{wohg64images} shows, the location of the ring is just
inside the side lobe of the dirty beam, and therefore we cannot entirely
exclude the possibility that it might be an artifact of the image
reconstruction inherent in the imperfect $u\varv$ coverage.
We refrain from further discussing
this structure until we can confirm it with better $u\varv$
coverage.

Figure~\ref{model2008}a shows the image predicted at 2.2~\mbox{$\mu$m}\ from
the dust torus model of Ohnaka et al. (\cite{ohnaka08}). The model image is 
characterized by the central star and the emission from the inner rim
on the far side (light red to white region in the southwest).
To examine whether the model
is consistent with the GRAVITY data, we show comparisons of the
visibility and closure phase in Figs.~\ref{model2008}b and \ref{model2008}c.
While the model closure phase is in fair agreement with the measurements,
the model predicts the 2.2~\mbox{$\mu$m}\ visibility to be too high compared to
the data. We computed 2.2~\mbox{$\mu$m}\ images at different inclination
angles within its uncertainty (0--30\degr) and different position angles
in the plane of the sky, but they all
show too high visibilities and too little elongation compared to the data. 
This is because the model predicts the stellar flux contribution at
2.2~\mbox{$\mu$m}\ to be too high compared to the GRAVITY data (note that
the central star's intensity is higher than the
intensity of the light red to white region in the southwest 
by a factor of at least $\sim$400). 

\begin{figure}
\begin{center}
%\centering
\resizebox{\hsize}{!}{\rotatebox{0}{\includegraphics{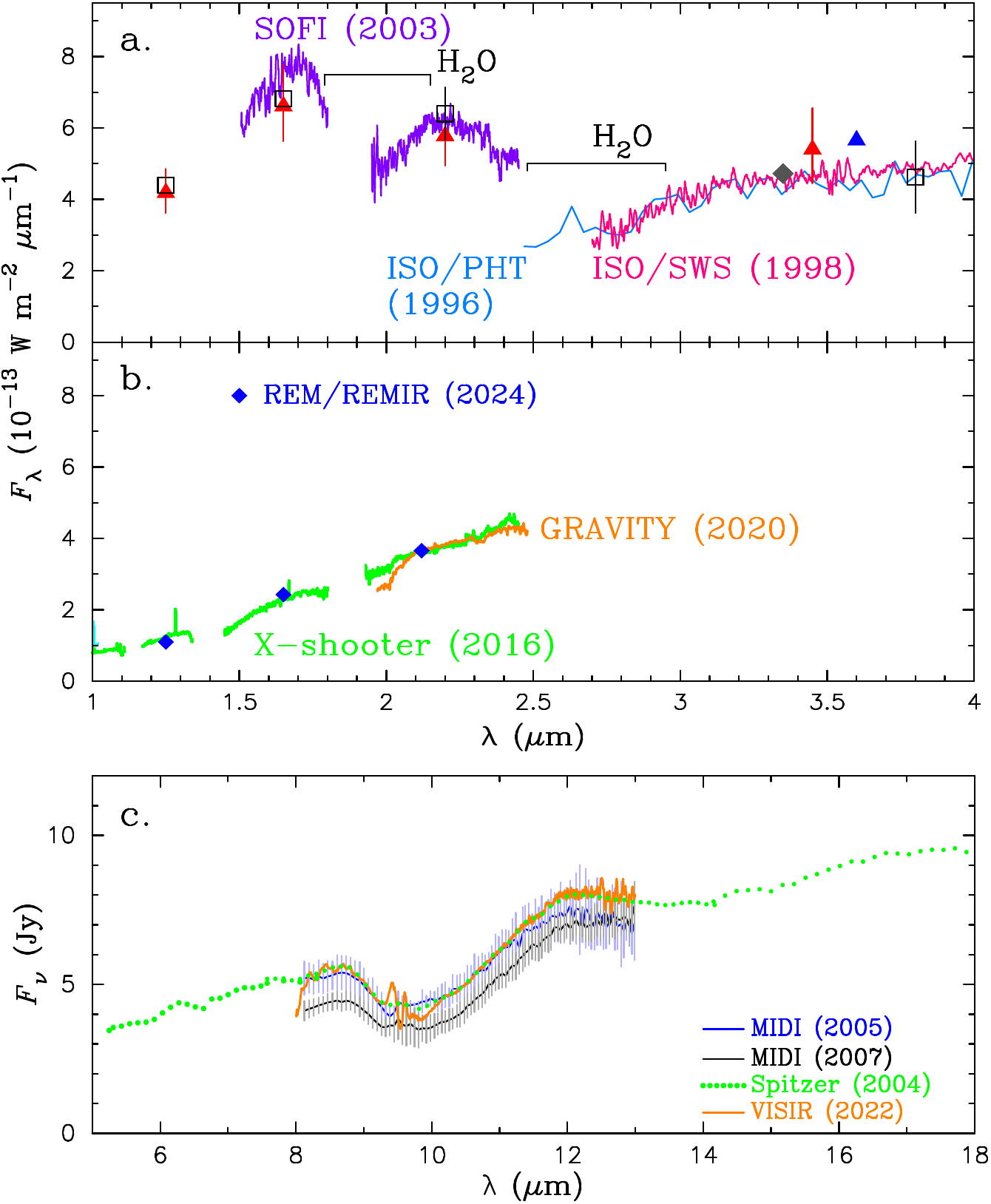}}}
%\resizebox{8.65cm}{!}{\rotatebox{0}{\includegraphics{specplot_1x3_nir_mir.pdf}}}
\end{center}
\caption{
Near- and mid-infrared spectrum of WOH G64.
{\bf a:} Spectrophotometric and photometric data taken at 1--4~\mbox{$\mu$m}\
before 2010. 
The light blue, pink, and violet lines represent the spectra taken with
ISO/PHT in 1996, ISO/SWS in 1998,
and SOFI in 2003, respectively. 
The red triangles and open squares correspond to the photometric data
from Whitelock et al. (\cite{whitelock03}) and Wood et al. (\cite{wood92}),
respectively. The bars show the (peak-to-peak) variability amplitudes,
not the measurement errors.
The gray diamond and blue triangle represent the photometric data taken
with WISE in 2009--2010 and \textit{Spitzer} in 2005-2006, respectively. 
The \mbox{H$_2$O}\ absorption features in the RSG spectrum are also
indicated.
{\bf b:} Spectra of WOH G64 taken more recently. The green and orange lines
represent the spectra taken with X-shooter in 2016 and GRAVITY in 2020,
respectively.
The blue diamonds show the {\it JHK$^{\prime}$} photometric data taken in 2024 
with REM/REMIR. The GRAVITY and X-shooter data are scaled to the
REM/REMIR $K^{\prime}$-band flux. 
The spikes in the X-shooter spectrum
as well as the very tiny peak at $\sim$2.16~$\mu$m in the GRAVITY spectrum
are the residual of the correction
for the absorption lines in the spectrum of the calibrator.
{\bf c:} Mid-infrared spectra of WOH G64 showing the 10~\mbox{$\mu$m}\ silicate
feature in absorption. 
The blue and black lines represent the spectra obtained with MIDI in
2005 and 2007, respectively, by Ohnaka et al. (\cite{ohnaka08}).
The dotted green line
shows the \textit{Spitzer}/IRS spectrum taken in 2004, while the orange line
corresponds to the VISIR spectrum taken in 2022.
The wavelength range between 9.3 and 9.8~\mbox{$\mu$m}\ is severely affected
by the telluric absorption.
}
\label{wohg64nirspec}
\end{figure}

\section{Change in the spectral shape in the infrared}
\label{sect_spec}

We found evidence that the difference in the 2.2~\mbox{$\mu$m}\ stellar flux
contribution could be due to a systematic change 
in the circumstellar environment of WOH G64
between the MIDI observations in 2005 and 2007 and our GRAVITY observations
in 2020. 
Figure~\ref{wohg64nirspec}a shows
(spectro)photometric data of WOH G64 taken before 2010 collected from the
literature (Wood et al. \cite{wood92}; Whitelock et al. \cite{whitelock03};
Srinivasan et al. \cite{srinivasan09}, \textit{Spitzer} Space Telescope; 
Cutri et al. \cite{cutri14}, Wide-field Infrared Survey Explorer
(WISE); 
Vandenbussche et al. \cite{vandenbussche02}, 
Infrared Space Observatory (ISO)/Short Wavelength Spectrometer (SWS); 
Trams et al. \cite{trams99}, ISO/Infrared Photo-polarimeter (PHT)).
We also derived
an $HK$-band spectrum from the archival SOFI data 
taken with a spectral resolution of 1000 
(the reduction of the SOFI data is described in 
Appendix~\ref{appendix_sofi}).
The spectrum obtained at 3.95--4.10~$\mu$m with ISAAC in 2001
(Matsuura et al. \cite{matsuura05}) is flat, without noticeable signatures
of the SiO bands\footnote{The absolute flux scale is uncertain by a factor of
two, which is why the ISAAC spectrum was not included in
Fig.~\ref{wohg64nirspec}}, consistent with the ISO/SWS data.
These (spectro)photometric data
show the spectrum of the RSG with \mbox{H$_2$O}\ absorption bands, as is
indicated in Fig.~\ref{wohg64nirspec}a. 

On the other hand, Fig.~\ref{wohg64nirspec}b shows near-infrared spectra of
WOH G64 taken more recently with VLT/X-shooter 
(Program ID: 097.D-0605(A), P.I.: S. Goldman;
see Appendix~\ref{appendix_xshooter} for details of the X-shooter data) 
and GRAVITY 
as well as the REM/REMIR {\it JHK$^{\prime}$} photometric data. 
Because the absolute flux calibration of the X-shooter and GRAVITY spectra is
uncertain, we tentatively scaled them to the REM/REMIR $K^{\prime}$-band flux
just to compare the spectral shape (the absolute flux
scale likely varied among the X-shooter, GRAVITY, and REM observations). 
The X-shooter, GRAVITY, and REM/REMIR data all indicate a monotonically rising
continuum. 
This is in marked contrast to the RSG spectrum seen in
Fig.~\ref{wohg64nirspec}a.
The \mbox{H$_2$O}\
absorption is also much less pronounced in the X-shooter spectrum compared
to the SOFI data. The molecular spectral features in cool variable
stars like WOH G64 change with the variability phase and cycle.
The bars of the {\it JHK} photometric data in Fig.~\ref{wohg64nirspec}a 
%(Wood et al. \cite{wood92} and Whitelock et al. \cite{whitelock03}) 
show the variability amplitudes measured from March 1995 to April 1998
by Whitelock et al. (\cite{whitelock03}). The variability amplitudes
measured from January 1987 to November 1991 by Wood et al. (\cite{wood92})
are similar\footnote{The variabilities in the $J$ and $H$ bands
measured by Wood et al. (\cite{wood92}) are not shown in
Fig.~\ref{wohg64nirspec}a, because they 
overlap with those measured by Whitelock et al. (\cite{whitelock03}). 
Only the $K$- and $L$-band variabilities are indicated with the bars.
}.
The rising continuum spectrum is difficult to
explain by the variability seen in the $J$, $H$, and $K$ bands.

To examine the spectral variation at longer wavelengths, we obtained
a mid-infrared spectrum of WOH G64 at 8--13~\mbox{$\mu$m}\ with VLTI/VISIR
at a spectral resolution of 300 (Program ID: 110.23RT.004, P.I.: K. Ohnaka;
see Appendix~\ref{appendix_visir} for details of the observation and
data reduction).
Figure~\ref{wohg64nirspec}c shows the VISIR spectrum obtained in October
2022 together with the MIDI data taken in 2005 and 2007 as well as 
the \textit{Spitzer}/InfraRed Spectrograph (IRS) spectrum taken in 2005 and presented
in Ohnaka et al. (\cite{ohnaka08}).
  The VISIR spectrum shows the 10~\mbox{$\mu$m}\ silicate
  feature in absorption, in agreement with the \textit{Spitzer}/IRS and 
 MIDI spectra within the errors. 
A comparison of the MIDI and \textit{Spitzer} spectra with other data in the literature
presented in Ohnaka et al. (\cite{ohnaka08}, their Fig.~3) indicates that
the long-term variation in the mid-infrared spectrum of WOH G64 is small.

\section{Discussion and conclusion}
\label{sect_discuss}

The results that the noticeable change in the spectral shape is seen in
the near-infrared but not in the mid-infrared may be explained by
hot dust forming close to the star.
Haubois et al. (\cite{haubois19}, \cite{haubois23}) detected dust clouds
forming at
$\sim$1.5~\mbox{$R_{\star}$}\ around the optically bright RSG Betelgeuse (which is much less 
dust-enshrouded than WOH G64) by near-infrared and visible polarimetry. 
They concluded that transparent grains such as \mbox{Al$_2$O$_3$}\ and
Fe-free silicates (\mbox{MgSiO$_3$}\ and \mbox{Mg$_2$SiO$_4$}) can explain the
observed data, although they favor the Fe-free silicates over \mbox{Al$_2$O$_3$}. 
These grains condense close to the star at $\sim$1500~K. 
Dynamical modeling of winds of asymptotic giant branch stars
shows that Fe-rich silicates, whose absorption efficiency is much higher
than the aforementioned species, condense onto the transparent
grains a little farther out (H\"ofner et al. \cite{hoefner22}). 
If this also occurs in RSGs like WOH G64, 
%If dust forms at a condensation
%temperature of $\sim$1500~K at a few stellar radii,
and the grains with mantles of Fe-rich silicates 
are optically thick in the near-infrared, it could account
for the monotonically rising continuum spectrum seen in the near-infrared
and the high obscuration of the central star.

The absence of a noticeable change in the mid-infrared spectrum suggests that
the hot dust should be confined in a region close to the star.
This is in qualitative agreement with the semimajor and semiminor axes of 
the elongated emission of $\sim$13 and $\sim$9~\mbox{$R_{\star}$}, respectively. 
The elongated emission may be due to a bipolar outflow along the axis of
the dust torus.
  Goldman et al. (\cite{goldman17}) interpreted the 1612~MHz OH maser
  emission of WOH G64 at an expansion velocity of 23.8~\mbox{km s$^{-1}$}\ as 
  spherical expansion or a bipolar outflow along our line of sight. 
Alternatively, the elongation may be caused by the
interaction with an unseen companion. The torus model of Ohnaka et al.
(\cite{ohnaka08}) is characterized by an inner radius of 15~\mbox{$R_{\star}$}\
(2.4~mas), in which the elongated emission with 13 $\times$ 9~\mbox{$R_{\star}$}\
(2 $\times$ 1.5~mas) can fit, although this cannot be taken as definitive
evidence for a companion. 
Levesque et al. (\cite{levesque09}) discussed the possibility of an unseen
hot companion and proposed spectroscopy in the blue to prove or disprove
the hypothesis. However, neither positive nor negative detections have been
reported in the literature. 

While the SOFI data taken in 2003 clearly shows the \mbox{H$_2$O}\ absorption in
the RSG spectrum, the X-shooter spectrum taken in 2016 shows
the rising continuum. Therefore, the spectral shape change in the
near-infrared should have started at some epoch between 2003 and 2016. 
Although no near-infrared spectra taken between 2003 and 2016 can be found 
in the literature, we can set an additional constraint on when the spectral
change started as follows. 
Levesque et al. (\cite{levesque09}) derived an extinction toward the central
star $A_V = 6.82 \approx \tau_V$
by fitting the visible spectrum of WOH G64
taken in December 2009. This optical depth in the visible translates into
$\tau_K = 0.62$ if we adopt the complex refractive index of the warm silicate
of Ossenkopf et al. (\cite{ossenkopf92}) with the grain size ($a$)
distribution,
$\propto a^{-3.5}$, between 0.005 and 0.1~\mbox{$\mu$m},\ as Ohnaka et al.
(\cite{ohnaka08}) assumed ($\tau_K$ is lower, 0.18, 
for a single grain size of 0.1~\mbox{$\mu$m}). With $\tau_K < 1$, the
central star should still have been visible in the $K$ band in December 2009.
Therefore, it is likely that the formation of hot dust started
at some epoch between December 2009 and 2016, and had not yet taken place
at the time of our MIDI observations in 2005 and 2007.

\begin{figure}
\begin{center}
%\centering
\resizebox{\hsize}{!}{\rotatebox{0}{\includegraphics{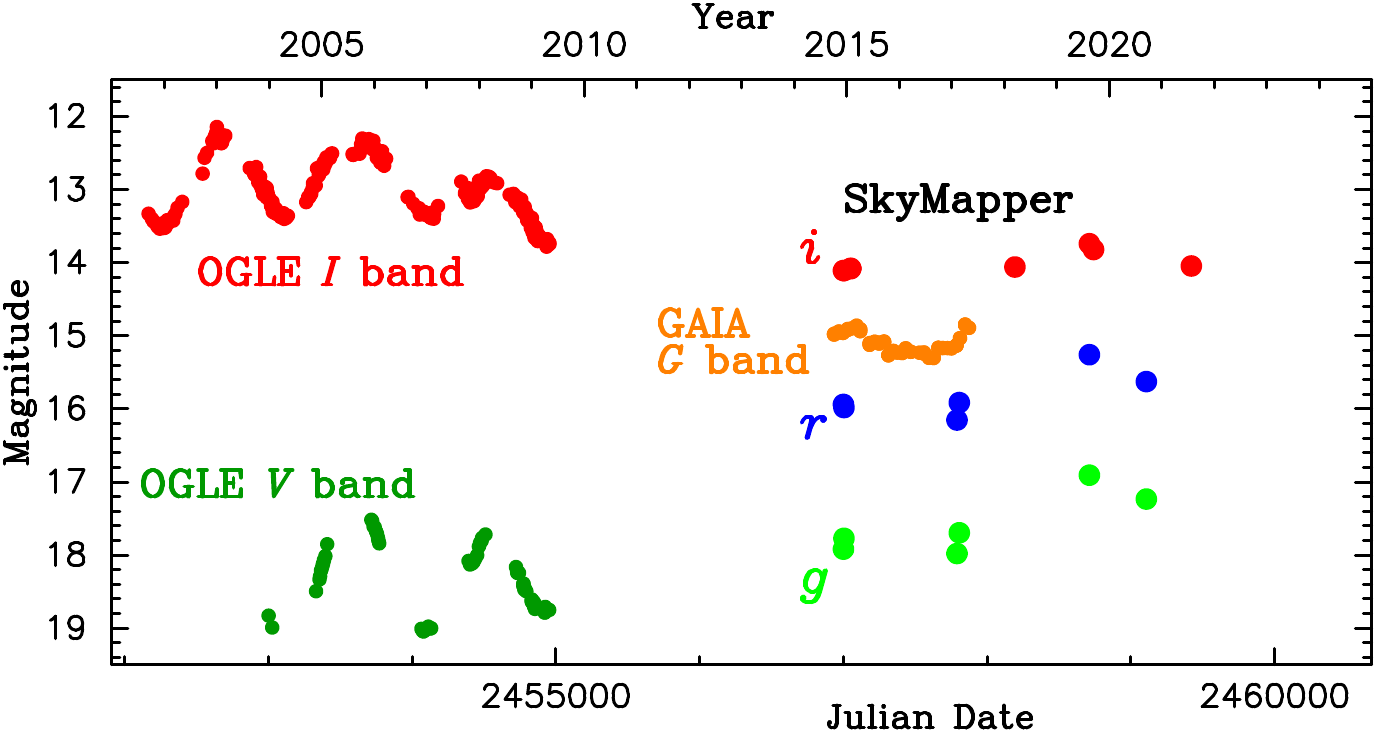}}}
%\resizebox{7cm}{!}{\rotatebox{0}{\includegraphics{wohg64_OGLE_GAIA_lightcurve3_NoREM.pdf}}}
\end{center}
\caption{
Visible light curves of WOH G64 from 2001 to 2021.
The red and dark green dots represent the photometric measurements in the
$I$ and $V$ bands, respectively, from the OGLE project.
The orange dots correspond to the \textit{Gaia} data in the $G$ band. 
The red, blue, and light green dots represent the photometric data
from the SkyMapper Southern Survey in the $i$, $r$, and $g$ bands,
respectively.
}
\label{wohg64lightcurve}
\end{figure}

The formation of new hot dust also means that the central star is now 
more obscured than the epochs before 2009. We collected visible photometric
data from the OGLE project (Soszy\'nski et al. \cite{soszynski09}) and
SkyMapper Southern Survey (Onken et al. \cite{onken24}) 
and show the visible light curves of WOH G64 in
Fig.~\ref{wohg64lightcurve}.
The $V$- and $I$-band light curves obtained by the OGLE project 
show the periodic variation until the middle of 2009. 
There are noticeable differences in the filter systems between OGLE and
SkyMapper. However, the OGLE $I$ band and SkyMapper $i$ band are relatively
close, with central wavelengths of 801~nm (OGLE $I$ band) and 779~nm (SkyMapper
$i$ band).
The SkyMapper $i$-band data suggest that WOH G64 is 
fainter after the end of 2014 than in the years covered by the OGLE data. 
Still, the current sparsity of the data in the literature -- particularly
between 2010
and 2015, exactly when the formation of hot dust likely started -- makes it
difficult to examine whether WOH G64 appeared much fainter at the time of
our GRAVITY observations in 2020 than in the past.

It is also worth noting that WOH G64 seems to be brighter
in the $g$ band than in the $V$ band. 
Given that the $g$ band is centered at 510~nm, shorter than the OGLE $V$ band
at 540~nm, and the visible flux of WOH G64 steeply decreases at shorter
wavelengths (van Loon et al. \cite{vanloon05}), we do not expect the $g$-band
flux to be higher than the $V$-band flux.
This high $g$-band flux may be due to the scattering by
the transparent grains forming close to the star, as was proposed above,
although this is not conclusive. 
Continuous long-term multiband photometric monitoring covering
at least the pulsation period of 840--930~days as well as visible
spectroscopic monitoring is necessary to confirm or
refute the effects of the new dust formation in the visible.

While the mid-infrared spectrum of WOH G64 shows little long-term change, 
it is interesting to probe whether there has been a
change in the structure of the circumstellar environment on spatial scales
larger than the new dust formation region close to the star. 
The VLTI/MATISSE instrument (Lopez et al. \cite{lopez22}) allows us to
obtain spectro-interferometric data in the $L$
(3--4.2~\mbox{$\mu$m}), $M$ (4.5--5.0~\mbox{$\mu$m}), and $N$ bands
(8--13~\mbox{$\mu$m}). MATISSE observations
and radiative transfer modeling will set much tighter constraints on its
circumstellar environment than MIDI did.

\begin{acknowledgement}
We thank the ESO Paranal team for supporting our VLTI observations and
the d'REM team for carrying out our REMIR observations. 
K.O. acknowledges the support of the Agencia Nacional de 
Investigaci\'on Cient\'ifica y Desarrollo (ANID) through
the FONDECYT Regular grant 1240301. 
This research made use of the \mbox{SIMBAD} database, 
operated at the CDS, Strasbourg, France. 
This publication makes use of data products from the Wide-field Infrared
Survey Explorer, which is a joint project of the University of California, Los
Angeles, and the Jet Propulsion Laboratory/California Institute of Technology,
funded by the National Aeronautics and Space Administration.
This research has made use of the NASA/IPAC Infrared Science Archive, which is
funded by the National Aeronautics and Space Administration and operated by
the California Institute of Technology.
This publication makes use of data products from the Two Micron All Sky
Survey, which is a joint project of the University of Massachusetts and the
Infrared Processing and Analysis Center/California Institute of Technology,
funded by the National Aeronautics and Space Administration and the National
Science Foundation.
This work has made use of data from the European Space Agency (ESA) mission
{\it Gaia} (https://www.cosmos.esa.int/gaia),
processed by the {\it Gaia} Data Processing and Analysis Consortium (DPAC,
https://www.cosmos.esa.int/web/gaia/dpac/consortium). Funding for the DPAC has
been provided by national institutions, in particular the institutions
participating in the {\it Gaia} Multilateral Agreement.
The national facility capability for SkyMapper has been funded through ARC
LIEF grant LE130100104 from the Australian Research Council, awarded to the
University of Sydney, the Australian National University, Swinburne University
of Technology, the University of Queensland, the University of Western
Australia, the University of Melbourne, Curtin University of Technology,
Monash University and the Australian Astronomical Observatory. SkyMapper is
owned and operated by The Australian National University's Research School of
Astronomy and Astrophysics. The survey data were processed and provided by the
SkyMapper Team at ANU. The SkyMapper node of the All-Sky Virtual Observatory
(ASVO) is hosted at the National Computational Infrastructure
(NCI). Development and support of the SkyMapper node of the ASVO has been
funded in part by Astronomy Australia Limited (AAL) and the Australian
Government through the Commonwealth's Education Investment Fund (EIF) and
National Collaborative Research Infrastructure Strategy (NCRIS), particularly
the National eResearch Collaboration Tools and Resources (NeCTAR) and the
Australian National Data Service Projects (ANDS).

\end{acknowledgement}

\begin{appendix}

\section{Observation log}
\label{appendix_obslog}

Table~\ref{obslog} shows the summary of our GRAVITY observations of
WOH G64.

\begin{table}[hbt]
\caption {
Our VLTI/GRAVITY observations of WOH G64.
}
\begin{center}
{
  \tabcolsep = 2pt
  \begin{tabular}{l c c c c c c}\hline
\# & $t_{\rm obs}$ & $B_{\rm p}$ & PA     & Seeing   & $\tau_0$ & N$_{\rm exp}$\\ 
   & UTC         &  (m)       & (\degr) & (\arcsec) &  (ms)  &  \\
\hline
\multicolumn{7}{c}{WOH G64}\\
\multicolumn{7}{c}{DIT = 30~s, N$_{\rm f}$ = 12}\\
\hline
\multicolumn{7}{c}{2020 December 15}\\
\multicolumn{7}{c}{AT configuration: A0-G1-J2-J3}\\
\hline

1 & 01:57:23 & 70.9/96.1/123.4/ & $-45$/$-81$/59/ & 0.57  & 6.2 & 1 \\
  &          & 56.3/126.1/79.6 & 52/26/8          &       &     &   \\
2 & 02:42:02 & 74.3/96.8/121.3/ & $-52$/$-87$/50/ & 0.43  & 6.6 & 2 \\
  &          & 55.5/127.7/82.4  & 43/16/$-2$      &   &         & \\ 
3 & 03:38:15 & 77.7/96.6/117.7/ & $-61$/85/39/    & 0.55  & 6.5 & 2 \\
  &          & 54.1/128.9/85.4  & 31/3/$-15$      &       &     &   \\
4 & 04:46:15 & 80.7/94.7/111.4/ & $-72$/75/25/    & 0.38  & 6.6 & 1 \\
  &          & 51.6/129.1/88.0  & 16/$-14$/$-30$  &       &     &   \\
5 & 05:39:33 & 81.9/92.0/105.4/ & $-81$/67/14/    & 0.31  & 6.5 & 1 \\
  &          & 49.2/128.2/89.2  & 4/$-26$/$-42$   &       &     &   \\
6 & 06:30:36 & 82.4/88.4/98.4/ & $-89$/60/2/     & 0.32  & 7.5 & 2 \\
   &          & 46.4/126.7/89.9 & $-8$/$-38$/$-53$ &      &     &   \\
7 & 07:30:54 & 82.0/83.0/89.8/ & 82/51/$-11$/     &  0.34  & 6.9 & 2 \\
   &          & 43.0/124.8/90.3 & $-23$/$-53$/$-67$ &       &     &   \\

\hline
\multicolumn{7}{c}{2020 December 26}\\
\multicolumn{7}{c}{AT configuration: A0-B2-C1-D0}\\
\hline

8 & 01:13:48 & 18.8/9.4/19.9/ & 77/77/8/    & 0.77  & 3.9 & 1 \\
  &          & 28.2/32.0/18.6  & 77/42/$-20$  &      &     &  \\
9 & 02:11:48 & 18.4/9.2/20.8 & 67/67/$-5$/ & 0.72 & 4.3 &  1 \\
  &          & 27.6/31.7/20.0 & 67/29/$-31$ &      &     &    \\
10 & 04:20:12 & 16.8/8.4/22.1/ & 45/45/$-34$ & 0.70 & 4.8 & 1 \\
   &          & 25.1/30.3/22.1 & 45/$-1$/$-56$ &    &     &   \\
11 & 05:04:03 & 15.9/7.9/22.3/ & 37/37/$-43$/ & 0.70  & 4.7 & 2 \\
  &          & 23.8/29.4/22.5 & 37/$-12$/$-64$ &      &     &  \\

\hline
\label{obslog}
\vspace*{-5mm}

\end{tabular}
}
\end{center}
\tablefoot{
$B_{\rm p}$: Projected baseline length.  PA: Position angle of the baseline 
vector projected onto the sky. 
DIT: Detector Integration Time.  $N_{\rm f}$: Number of frames in each 
exposure.  $N_{\rm exp}$: Number of exposures. 
The seeing and the coherence time ($\tau_0$) were measured in the visible.
}
\end{table}

%\FloatBarrier

\section{Observed visibilities and closure phases of WOH G64}
\label{appendix_vis_wl}

\begin{figure}
\begin{center}
\resizebox{0.97\hsize}{!}{\rotatebox{0}{\includegraphics{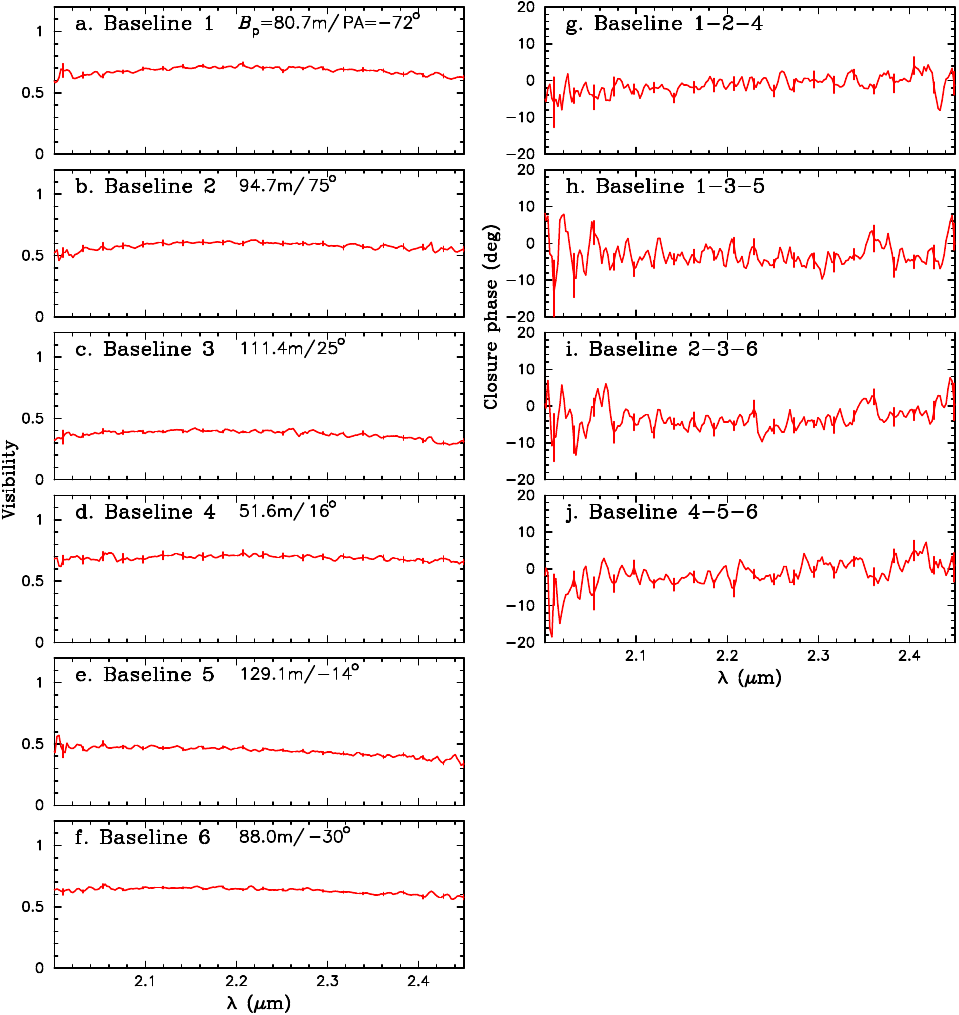}}}
\end{center}
\caption{
Visibilities and closure phases of WOH G64 observed at the A0-G1-J2-J3
configuration. 
{\bf a--f:} Visibility. The projected baselines ($B_{\rm p}$) and position
angle (PA) are given in each panel. 
{\bf g--j:} Closure phase. The baselines that form the telescope triplet
are given in each panel. 
}
\label{viscpplotL}
\end{figure}

\begin{figure}
\begin{center}
\resizebox{0.97\hsize}{!}{\rotatebox{0}{\includegraphics{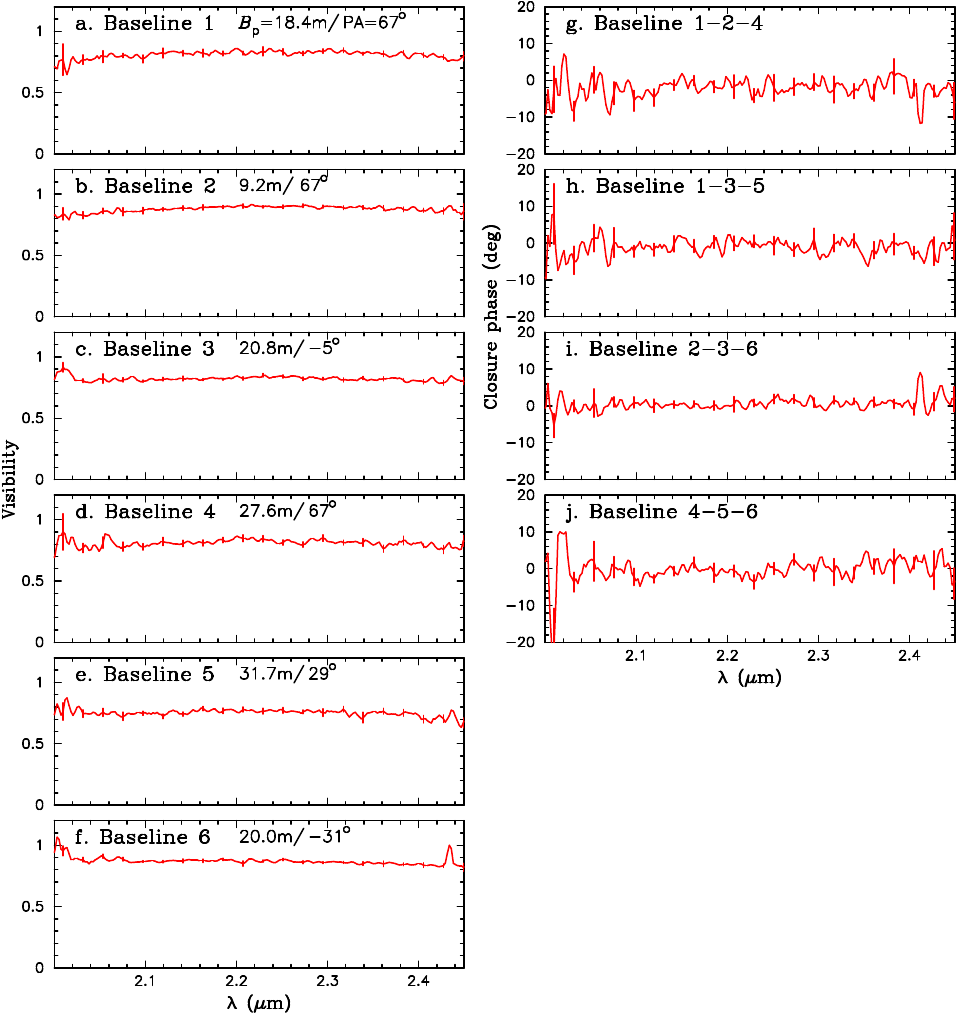}}}
\end{center}
\caption{
Visibilities and closure phases of WOH G64 observed at the A0-B2-C1-D0
configuration, shown in the same manner as Fig.~\ref{viscpplotL}.
}
\label{viscpplotS}
\end{figure}

Figures~\ref{viscpplotL} and \ref{viscpplotS} show examples of the 
visibilities and closure phases observed at the A0-G1-J2-J3 and 
A0-B2-C1-D0 configurations, respectively. The observed interferometric
data (spectrally binned to the resolution of 330) 
do not show signatures of the CO bands longward of 2.3~$\mu$m.

\FloatBarrier
%\clearpage

\section{Calibration of the GRAVITY spectra}
\label{appendix_gravity_spec}

We obtained the spectroscopically calibrated spectrum of WOH G64 
as follows: 
\[
F_{\rm sci}^{\rm true} = \frac{\eta_{\rm cal}}{\eta_{\rm sci}} F_{\rm
  sci}^{\rm obs} \times \frac{F_{\rm cal}^{\rm true}}{F_{\rm cal}^{\rm obs}},
\]
where $F_{\rm sci (cal)}^{\rm true}$ and $F_{\rm sci (cal)}^{\rm obs}$
denote the true and observed spectra of the science target (sci) or 
the calibrator (cal), respectively. $\eta_{\rm sci (cal)}$ represents the
fraction of the flux injected into the fibers of the beam combiner
for the science target or calibrator. 
The spectrum was derived from the data without the spectral binning, because
the S/N of the spectral data was sufficiently high. 
We used the calibrator HD37379 (F6/7V) for the spectrophotometric
calibration. 
To approximate the true spectrum of HD37379, we used the flux-calibrated 
spectrum of
HD126660 (F7V) taken with the InfraRed Telescope Facility (IRTF Spectral
Library\footnote{http://irtfweb.ifa.hawaii.edu/\~{}spex/IRTF\_Spectral\_Library/}, Rayner et al. \cite{rayner09}) because its spectral type and
luminosity class are very close to those of HD37379.
The IRTF spectrum taken with a spectral resolution of 
$\lambda/\Delta \lambda = 2000$ was convolved to match the spectral
resolution of 500 of our GRAVITY (spectral) data and then scaled to match the
$K$ magnitude of HD37379 by multiplying by
$f_{K, {\rm HD37379}}/f_{K, {\rm HD126660}}$, where
$f_{K, {\rm HD37379\, (or\, HD126660)}}$ denotes the $K$-band flux of HD37379 
or HD126660. The spectrum obtained in this manner was used for 
$F_{\rm cal}^{\rm true}$ in the above spectroscopic calibration. 

While the spectra obtained with four ATs agree well in the shape, the
absolute flux calibration of the GRAVITY spectra is difficult. 
First, 
the fraction of the flux injected into the GRAVITY's beam combiner fibers
depends on the performance of the adaptive optics (AO) system NAOMI of the
ATs (Woillez et al. \cite{woillez19}). While the calibrator HD37379 is
bright enough for NAOMI to work properly, the $G$-band magnitude of
WOH G64 is just at the limit of NAOMI ($G$ = 15), which resulted in
degraded AO performance for WOH G64 compared to HD37379. This means that
the fraction of the flux injected into the fibers was systematically
lower for WOH G64 than for HD37379. 
  Furthermore, the fiber injection fraction also depends on other factors
  such as the aberration and tip/tilt correction.
  Jovanovic et al. (\cite{jovanovic17}) show that the coupling efficiency
  to single-mode fibers varies up to 30\%. 
  Given this large uncertainty, we tentatively scaled the GRAVITY spectra
  to the $K^{\prime}$-band magnitude measured with REM/REMIR.

%\FloatBarrier

\section{REM observations of WOH G64}
\label{appendix_rem}

The RSG WOH G64 was observed with the REM/REMIR camera with the $J$, $H$, and 
$K^{\prime}$
filters on 2024 August 11 (UTC). We obtained 10 frames with each filter,
using an exposure time of 15~s for each filter. After flat-fielding, 
sky subtraction, and averaging of the frames, the flux of WOH G64 was
measured by PSF (point spread function) photometry and calibrated
using five stars in the same field of view, for which the 2MASS
{\it JHK$_s$} magnitudes are available.
The 2MASS $K_s$ magnitudes were converted to the $K^{\prime}$ magnitudes
using the relations given on the GEMINI web
page\footnote{https://www.gemini.edu/observing/resources/near-ir-resources\\/photometry/niri-filter-color-transformations}
and the 2MASS web page\footnote{https://irsa.ipac.caltech.edu/data/2MASS/docs/releases/allsky/doc/\\sec6\_4b.html}. 
The $J$-, $H$-, and $K^{\prime}$-band fluxes of WOH G64 and their errors are
listed in Table.~\ref{rem_res}.

\begin{table}
\caption {
Near-infrared flux of WOH G64 measured with REM/REMIR on 2024 August 11.
}
\begin{center}

\begin{tabular}{l c}\hline
Band & Flux ($10^{-13}$ W~m$^{-2}$~$\mu$m$^{-1}$)  \\ 
\hline
$J$ & $1.10 \pm 0.04$ \\
$H$ & $2.43 \pm 0.07$  \\
$K^{\prime}$ & $3.66 \pm 0.19$ \\
\hline
\label{rem_res}
\vspace*{-5mm}

\end{tabular}
\end{center}
%\tablefoot{
%}
\end{table}

\FloatBarrier
%\clearpage

\section{Image reconstruction}
\label{appendix_gravity_viscp}

\begin{figure}[hbt]
\begin{center}
\resizebox{0.97\hsize}{!}{\rotatebox{0}{\includegraphics{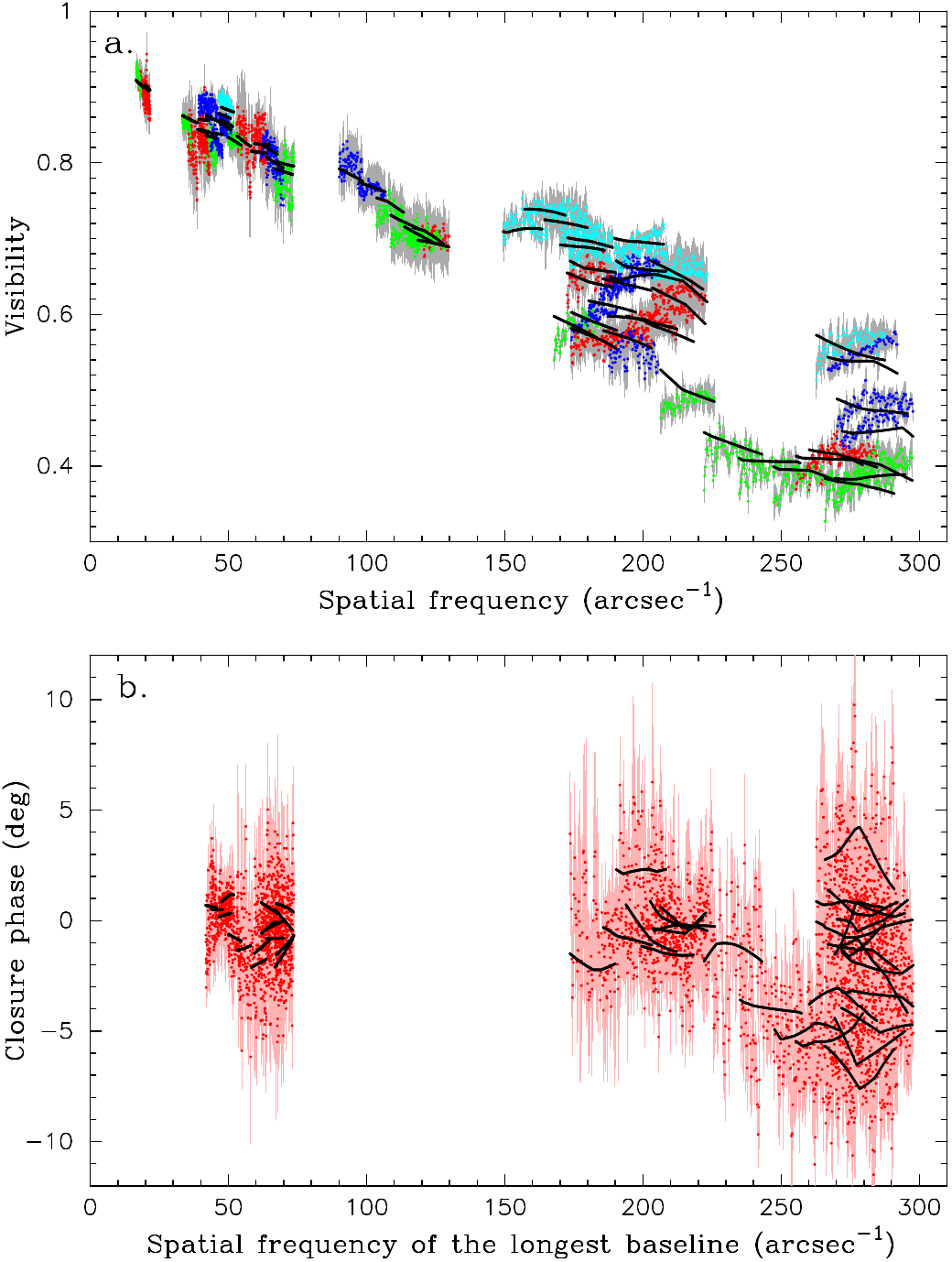}}}
\end{center}
\caption{
Comparison of the visibilities and closure phases of the reconstructed
image shown in Fig.~\ref{uv_image}b with the GRAVITY measurements.
{\bf a:} Visibility.
{\bf b:} Closure phase.
%{\bf
  In either panel,
  the black dots (they appear to be solid lines due to the high density) 
  represent the visibilities or closure phases computed from the
  reconstructed image. 
  The measurements are shown in the same manner as in
  Figs.~\ref{model2008}b and \ref{model2008}c.
%}
}
\label{gravity_viscpplot}
\end{figure}

Comparisons of the visibilities and closure phases computed from the IRBis
reconstructed image and the GRAVITY data are shown in
Fig.~\ref{gravity_viscpplot}. 

\begin{figure*}[t!]
%\begin{center}
\centering
\resizebox{\hsize}{!}{\rotatebox{-90}{\includegraphics{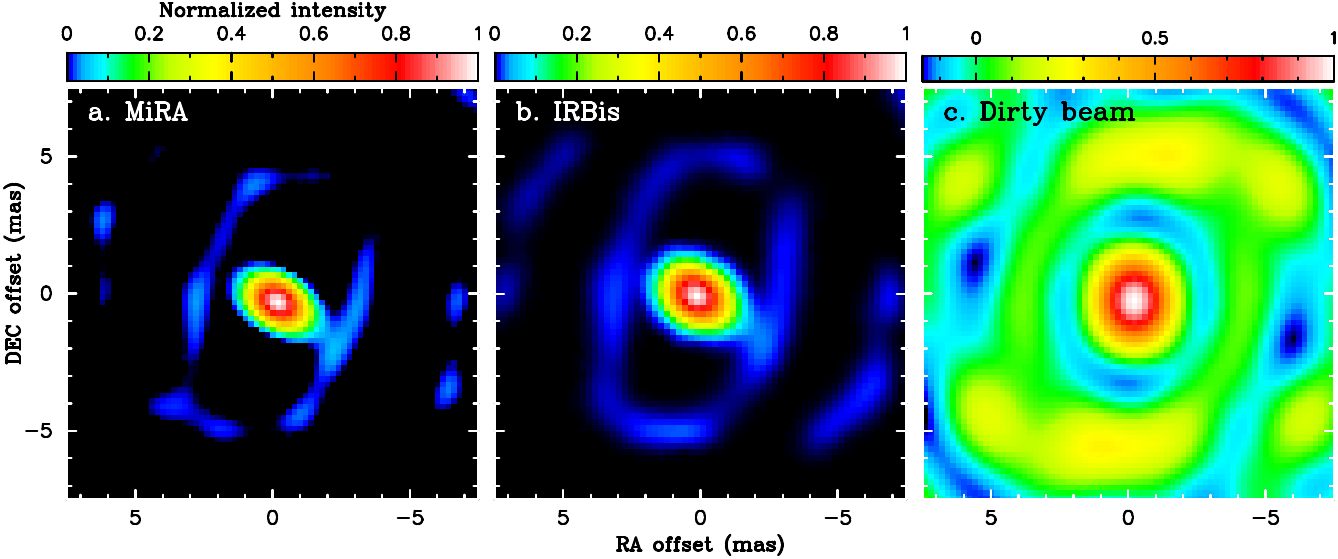}}}
%\end{center}
\caption{
Images of WOH G64 reconstructed at 2.2~\mbox{$\mu$m}\ with MiRA and IRBis.
{\bf a:} Image reconstructed using MiRA with the pixel difference
quadratic regularization. 
{\bf b:} Image reconstructed with IRBis (the same image as in
Fig.~\ref{uv_image}b).
{\bf c:} Dirty beam.
North is up, east to the left in all panels.
}
\label{wohg64images}
\end{figure*}

The image reconstructed with MiRA using the pixel difference quadratic
regularization  is shown in
Fig.~\ref{wohg64images}, together with the image obtained with \mbox{IRBis}
and the dirty beam. The MiRA image reconstructed with the pixel intensity
quadratic regularization is very similar to the one obtained with the
pixel difference quadratic regularization. A flat prior was used in both
reconstructions with MiRA. 

  We reconstructed images using the data with spectral windows centered at
  2.2~\mbox{$\mu$m}\ with different widths of 0.05, 0.1, and 0.2~\mbox{$\mu$m}. 
  The reconstructed images appear to be very similar. Therefore, we show
  the image obtained with the width of 0.2~\mbox{$\mu$m}, which slightly
  enhances the $u\varv$ coverage.

%\onecolumn
%\onecolumngrid
%\begin{multicols}{3}

%\end{multicols}
%\twocolumn

\FloatBarrier
%\clearpage

%\begin{multicols}{2}

\section{SOFI observations of WOH G64}
\label{appendix_sofi}

We downloaded the SOFI archival data of WOH G64 and the spectroscopic
calibrator HIP21984 (G3V) taken on 2003 June 29, covering from 1.5 to
2.5~\mbox{$\mu$m}\
with a spectral resolution of 1000 (Program ID: 71.B-0558(A),
P.I.: J. Blommaert). The data were processed with the
SOFI pipeline ver. 1.5.12. The spectroscopic calibration of the data of
WOH G64 was done in the same manner as described in
Appendix~\ref{appendix_gravity_spec} 
but with both $\eta_{\rm sci}$ and $\eta_{\rm cal}$ set to 1 because
they are irrelevant for the SOFI observations.
We used HD10697 (G3Va) as a proxy star because of its similarity to HIP21984 
with respect to the spectral type and luminosity class. 
The photometrically calibrated spectrum of HD10697 available in the IRTF
Spectral Library (Rayner et al. \cite{rayner09}) was convolved to match the
spectral resolution of 1000 of the SOFI data and then multiplied
by the $K$-band flux ratio $f_{K,\,{\rm HIP21984}}/f_{K,\,{\rm HD10697}}$.
This was used for 
$F_{\rm cal}^{\rm true}$ in the spectrophotometric calibration. 

%\FloatBarrier

\section{X-shooter observations of WOH G64}
\label{appendix_xshooter}

We downloaded the reduced (phase 3) X-shooter spectra of
WOH G64 and the spectroscopic calibrator HIP28322 (B8V) taken on 2016 July 27
(Program ID: 097.D-0605(A), P.I.: S. Goldman).
While the reduced data of WOH G64 are
spectroscopically calibrated, the telluric lines are not removed.
We attempted to remove them as much as possible in the same manner as 
described in
  Appendix \ref{appendix_gravity_spec} with 
  both $\eta_{\rm sci}$ and $\eta_{\rm cal}$ set to 1.
To obtain $F_{\rm cal}^{\rm true}$, we used HD147550 (B9V) as a proxy star.
Its near-infrared spectrum available in the X-shooter Spectral Library
(Gonneau et al. \cite{gonneau20}),
which is flux-calibrated and telluric-corrected,
was scaled to match the $K$-band flux of HIP28322 and used for
$F_{\rm cal}^{\rm true}$ in the spectroscopic calibration.
The resulting spectrum was convolved to match the spectral resolution of
the GRAVITY spectrum of WOH G64. 
However, as mentioned in Gonneau et al. (\cite{gonneau20}),
the absolute flux calibration of the X-shooter spectrum is difficult due to
the slit loss. Therefore, we tentatively scaled the X-shooter spectrum to
  match the $K^{\prime}$-band flux measured with REM/REMIR.

\section{VISIR observation of WOH G64}
\label{appendix_visir}

Our VISIR observation of WOH G64 at 8--13~\mbox{$\mu$m}\ took place on
2022 October 7 (UTC). A slit width of 1\arcsec\ was used, which resulted in
a spectral resolution of 300. 
The data of WOH G64 and the calibrator HD33554 
were first reduced with the VISIR pipeline ver. 4.4.2, and
the spectrum of WOH G64 was calibrated in the same manner as described
in Appendix~\ref{appendix_gravity_spec} with both $\eta_{\rm sci}$ and
$\eta_{\rm cal}$ set to 1. 
We used the absolutely calibrated spectrum of HD33554 presented
in Cohen et al. (\cite{cohen99}) as $F_{\rm cal}^{\rm true}$.

%\end{multicols}
%\FloatBarrier

\end{appendix}

\end{document}